# Identifying Ketamine Responses in Treatment-Resistant Depression Using a Wearable Forehead EEG

Zehong Cao, *Member, IEEE*, Chin-Teng Lin\*, *Fellow, IEEE*, Weiping Ding, *Member, IEEE*,
Mu-Hong Chen, Cheng-Ta Li, Tung-Ping Su\*

*Abstract* - This study explores responses to ketamine in patients with treatment-resistant depression (TRD) using a wearable forehead electroencephalography (EEG) device. We recruited and randomly assigned 55 outpatients with TRD into three approximately equal-sized groups (A: 0.5 mg/kg ketamine; B: 0.2 mg/kg ketamine; and C: normal saline) under double-blind conditions. The ketamine responses were measured by EEG signals and Hamilton Depression Rating Scale (HDRS) scores. At baseline, the responders showed significantly weaker EEG theta power than the non-responders ($p < 0.05$). Compared to baseline, the responders exhibited higher EEG alpha power but lower EEG alpha asymmetry and theta cordance post-treatment ($p < 0.05$). Furthermore, our baseline EEG predictor classified the responders and non-responders with 81.3 ± 9.5% accuracy, 82.1 ± 8.6% sensitivity and 91.9 ± 7.4% specificity. In conclusion, the rapid antidepressant effects of mixed doses of ketamine are associated with prefrontal EEG power, asymmetry and cordance at baseline and early post-treatment changes. Prefrontal EEG patterns at baseline may serve as indicators of ketamine effects. Our randomized, double-blind, placebo-controlled study provides information regarding the clinical impacts on the potential targets underlying baseline identification and early changes from the effects of ketamine in patients with TRD.

This work was supported by the Australian Research Council (ARC) [DP180100670, DP180100656] and the Army Research Laboratory with Cooperative Agreement [W911NF-10-2-0022]. The research was also partially sponsored by the Ministry of Science and Technology of Taiwan [MOST 106-2218-E-009 -027 -MY3, MOST 106-2221-E-009 -016 -MY2]. Additionally, the study was supported by grants from the National Scientific Council of Taiwan [NSC 101-2314-B-010-060, 102-2314-B-010-060] and the Brain Research Center of National Yang-Ming University, Taiwan.

Z. Cao and C.T. Lin are affiliated with the Center for Artificial Intelligence and Faculty of Engineering and Information Technology, University of Technology Sydney, Australia (*correspondence to Chin-Teng.Lin@uts.edu.au).

W. Ding is affiliated with the Faculty of Engineering and Information Technology, University of Technology Sydney, Australia and the School of Computer Science and Technology, Nantong University, Nantong, China.

M.H. Chen and C.T. Li are affiliated with Taipei Veterans General Hospital, Taipei, Taiwan.

T.P Su is affiliated with Cheng-Hsin General Hospital, Taipei, Taiwan, and the Department of Psychiatry, Taipei Veterans General Hospital, Taipei, Taiwan (*correspondence to tomsu0402@gmail.com).

## I. INTRODUCTION[1]

DEPRESSION is a common affective disorder with a high prevalence worldwide (4.5% - 37.4%), and the incidence of major depressive disorder ranges from 4.6% to 9.3%. Current traditional antidepressants, including selective serotonin reuptake inhibitors, serotonin norepinephrine reuptake inhibitors and norepinephrine and dopamine reuptake inhibitors, have been used to treat depression individually or in combination. Approximately 30-40% of patients with major depression do not respond to traditional antidepressants, and a patient with major depression who does not respond to at least two clinical trials of traditional antidepressants during an adequate period is considered to have treatment-resistant depression (TRD) [1]. Patients with TRD have traditionally been defined as patients with no response or an inadequate response; however, recent clinical studies have shown significant improvements in depressive symptoms after a low-dose ketamine treatment. Ketamine, which is an NMDA receptor antagonist, differs from traditional antidepressants, which all act on monoamine systems. In contrast, ketamine acts on the glutamate system in the brain. The discovery and replication of the rapid and robust antidepressant effects of ketamine on TRD symptoms may represent an important achievement, addressing some limitations of current antidepressant treatments, such as a longer latency to become effective. Additionally, low-dose ketamine has been noted to have robust and rapid antidepressant effects, specifically in patients with TRD [2-4].

Depressive symptoms are associated with brain cortical abnormalities, and the prefrontal cortex (PFC) is critically involved in the neuro-circuitry of depression [5]. Depressive disorders are correlated with a reduction in dorsolateral PFC gray matter volumes and unique directional changes in the prefrontal cortex [6, 7]. Furthermore, as shown in our previous studies involving patients with TRD, glucose metabolism in the PFC is correlated with ketamine responses [4], and these patients continued to present abnormal glucose metabolism in the PFC after aggressive treatments [8, 9]. Therefore, the PFC is a potential indicator reflecting the neurophysiological activity of patients with TRD.

Electroencephalography (EEG), which features non-invasive electrodes placed along the scalp, is an electrophysiological monitoring method used to record spontaneous electrical activity in the brain [10]. EEG has been widely used to study antidepressant treatment responses in many clinics due to its broad availability and cost-effectiveness [11, 12]. The development of dry sensors has resulted in wearable EEG headbands with dry electrodes that are more convenient for measuring EEG signals than conventional EEG headsets with wet electrodes. Dry electrodes avoid the use of conductive gel and skin preparation while transmitting brainwaves that are encountered with wet electrodes [13, 14].

The relationship between the antidepressant response and prefrontal EEG dynamics, particularly measurements derived from theta and alpha activity, has attracted significant interest [12, 15-17]. Alpha activity has an inhibitory effect on cortical network activity; thus, increased alpha power has been observed in depressed patients who respond to antidepressant treatment [18]. Some studies have also reported antidepressant treatment responses in association with hemispheric asymmetry, which is a relative measure of the difference in EEG alpha power between the right and left prefrontal regions that reflects relatively lateralized cortical activity [19]. While left versus right hemispheric activation has been associated with treatment responses, a body of literature has raised the possibility of hemispheric asymmetry in the alpha band [11, 18, 20, 21]. Similarly, conflicting studies have investigated theta activity and reported that increased absolute theta power [22] or decreased relative theta power [17] is associated with the treatment response. To determine the emergent effectiveness of antidepressant therapy, theta cordance [23] has been derived from both absolute and relative theta power. Prior studies have shown decreased frontal theta cordance after antidepressant treatment [24, 25]. Additionally, "pre-exposure EEG" may capture the trait aspects of psychological dysfunction [26], suggesting that a baseline (pre-treatment) EEG has the potential to predict the effects of antidepressants in an individual. These EEG patterns, which are correlated with the response to antidepressants, potentially represent the neurophysiological activities in patients with TRD.

According to a literature review, the derived EEG features demonstrate various trends from the baseline to post-treatment that may be associated with different types of antidepressants. However, whether the responses of TRD patients to ketamine involve the dynamics of prefrontal EEG and whether discriminative EEG can be used as a potential feature to predict the ketamine response remain unclear. The novelty of this study involves the dynamics of prefrontal EEG to investigate the responses of patients with TRD to ketamine treatment instead of traditional antidepressants. Because antidepressants exert their effects on PFC activation and EEG dynamics, the a priori hypothesis is that the rapid antidepressant effects of different doses of ketamine are correlated with baseline and early changes in EEG patterns in the prefrontal area similar to the time-dependent effects of some traditional antidepressants. Consistent with previous studies [27-29], our study proposes to investigate prefrontal EEG patterns (e.g., power, asymmetry and cordance) at the baseline level (-10 to 0 min) and early post-treatment changes (240-250 min after drug infusion) to determine the signatures of the ketamine response under randomized, double-blind, placebo-controlled conditions. Additionally, the discriminative EEG patterns at baseline may help TRD patients understand ketamine's effects and could potentially be used in daily life via a wearable forehead EEG device. To the best of our knowledge, no studies have investigated brain dynamics to identify ketamine responses in TRD patients using a wearable forehead EEG; thus, this novel study could contribute to biomedical engineering and clinical applications. In this study, we used a wearable EEG system and advanced data analysis approaches to investigate TRD patients using ketamine as an antidepressant.

This paper is organized as follows. First, Section II introduces the participant recruitment, experimental procedures, data processing and statistical analysis used in this study. Section III presents the experimental results, including the demographic characteristics and clinical and treatment responses, comparisons between groups, and predictor performance. Section V discusses the experimental results, future directions and limitations. Section VI provides the conclusion.

## II. MATERIALS AND METHODS

### A. Subjects

Outpatients with depression were recruited at the Psychiatric Department of Taipei Veterans General Hospital (VGH). The patients' diagnoses were based on the criteria from the Diagnostic and Statistical Manual of Mental Disorders IV Text Revision (DSM-IV-TR) for the diagnosis of major depressive disorder. In addition, the included patients did not have major medical or neurological illnesses. The enrolled patients were required to have a poor treatment history with at least two different antidepressants (with sufficient dosages and treatment periods) and at least one invalid effort with adequate antidepressant treatment during their current depressive episode to ensure that each subject was resistant to antidepressant medications. Patients were excluded if they received a structured diagnostic Mini International Neuropsychiatric Interview (MINI) [30] conducted by a board-certified psychiatrist (TPS) to confirm the diagnosis.

The Hamilton Depression Rating Scale-17 (HDRS-17) [31] is the most widely used clinician-administered depression assessment scale for evaluating antidepressants' effects. This questionnaire, which was designed for adults, is used to rate the severity of patients' depression by probing their mood, feelings of guilt, suicidal ideation, insomnia, agitation or retardation, anxiety, weight loss and somatic symptoms. The original version contains 17 items pertaining to symptoms of depression experienced over the previous week. A score of 0–7 is considered normal, while a score of 20 or higher (indicating at least moderate severity) is usually required for entry into a clinical trial.

Patients with a history of bipolar disorder, psychotic depression or substance/alcohol abuse or a score of less than 18 on the HDRS-17 at screening and less than 13 at the start of infusion were excluded from the study. Patients with comorbid Axis I diagnoses of anxiety disorders were permitted due to the ubiquity of these disorders among patients with major depressive disorder (MDD). This study was approved by the Institutional Review Board of the Taipei VGH (2012-04-037B), and informed consent was obtained from all subjects before entry into the study. This clinical trial was registered with the UMIN Clinical Trials Registry (UMIN000016985) on June 30, 2016. Additionally, all experimental procedures were performed in accordance with the relevant guidelines and regulations.

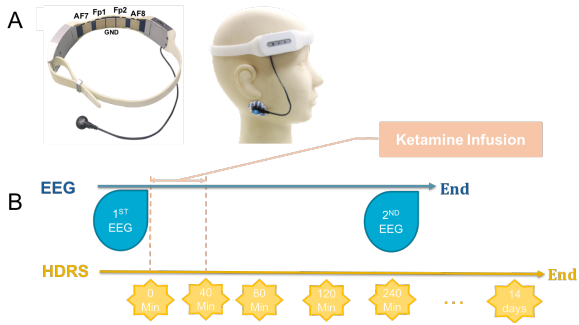

Figure 1. EEG analytical procedures used in the ketamine experiment. (A) Wearable EEG device; (B) Ketamine infusion and EEG and HDRS recordings.

### B. Experimental Procedures

Each patient received a clinical assessment, including a psychiatric profile, medical history and assessment of medication use. The patients with TRD were divided into the following three subgroups at an approximate 1:1:1 ratio: group A-0.5 mg/kg ketamine, group B-0.2 mg/kg ketamine, and group C-normal saline (NS). This subgroup allocation was conducted under randomized and double-blind conditions by an independent nurse who was not involved in our study. The experiments were performed in a quiet, dimly lit room at psychiatric inpatient unit D020 of Taipei VGH. During the first two minutes of the experiment, the patients were instructed to take several deep breaths while they adapted to the environment. A specific wearable EEG recording system (Fig. 1A) called the "Mindo-4S Jellyfish" (Brain Rhythm Inc., Zhubei District, Hsinchu, Taiwan) that used a 16-bit quantization level at a sampling rate of 512 Hz was used in this study; the system included four dry electrodes (Fp1, Fp2, AF7 and AF8) in the prefrontal region, a reference electrode (A2) [13, 14] and a built-in real-time EEG signal enhancement to remove artifacts [32]. These dry electrodes are more convenient for measuring EEG signals than conventional wet electrodes and are preferred because they avoid the use of conductive gel and skin preparation while achieving a signal quality comparable to that of wet electrodes [13, 14].

Specifically, a real-time artifact removal algorithm based on canonical correlation analysis (CCA), feature extraction, and a Gaussian mixture model (GMM) were used to improve the quality of the EEG signals. The CCA was used to decompose the EEG signals into components, followed by feature extraction to extract the representative features and GMM to cluster these features into groups to recognize and remove artifacts. The feasibility of the proposed algorithm was demonstrated by effectively removing artifacts caused by blinks, head/body movement, and chewing from the EEG recordings. Furthermore, the dry EEG sensors enabled the continuous monitoring of the high-temporal resolution brain dynamics without requiring the application of conductive gels on the scalp. The long-term impedance measurements suggest that compared to other gel electrode types, the dry EEG sensors have the potential to provide stable EEG signals. Notably, our developed mobile and wireless EEG system features a dry micro-electromechanical system with electroencephalography sensors, low-power signal acquisition, amplification and digitization, wireless telemetry, online artifact enhancement, and signal pre-processing. The test of our EEG system showed promising and consistent EEG signal qualities for all subjects [13, 14].

Then, as shown in Fig. 1B, the patients were instructed to close their eyes as the EEG signals were recorded for approximately 10 min (baseline, 1st session). Subsequently, an infusion of 0.5 mg/kg ketamine, 0.2 mg/kg ketamine or NS was administered over the course of 40 min. Notably, the patients always experienced difficulty in sitting still or had a headache during the ketamine administration; therefore, the EEG signals were not recorded during this period. The subjects were also allowed to sit in the rest area and read books to avoid fatigue or drowsiness. Subsequently, closed-eye EEG signals were recorded during a second 10-min session to evaluate the effects of ketamine 240-min post-infusion. Additionally, an experienced EEG specialist monitored the experimental conditions during the EEG recordings (e.g., ensuring that the location of the wearable EEG device did not shift).

In addition, the quantitative depressive symptom ratings were recorded 0 min (baseline), 40 min (end of infusion), 80 min, 120 min, 240 min, 24 hours (on day 2), 3 days, 4 days, 5 days, 6 days, 7 days (one week) and 14 days (two weeks) post-treatment using the HDRS-17 (Fig. 1B). A "responder" was defined as an individual exhibiting at least a 45% reduction in the HDRS-17 score from baseline to 240-min post-treatment, and all other individuals were classified as non-responders.

### C. EEG Data Processing

The EEG data were analyzed with EEGLAB, which is an open-source MATLAB toolbox for electrophysiological signal processing and analysis [33]. The analytical procedures used to process the EEG signals included bandpass filtering, time-frequency analysis, power estimation, asymmetry calculation, cordance assessment and antidepressant treatment response (ATR) measurement.

Experienced EEG specialists were required to inspect the collected data to ensure that the raw EEG data did not have interference from artifacts despite the fact that the wearable EEG device has a built-in function to remove artifacts via real-time EEG signal enhancement. Then, the EEG data were bandpass-filtered from 1 to 12 Hz using a zero-phase finite impulse response filter. The processed EEG signals were subjected to further analysis. The main aim of the data processing approaches is to extract correct information from the brain, and then, the extracted features can classify

**Table I. Demographics, Clinical Profiles and Treatment Responses of the Ketamine and NS Groups**

| Characteristics | Antidepressant Groups | | | Bonferroni-adjusted one-way ANOVA p |
|---|---|---|---|---|
| | Ketamine (n = 37) | | C: NS (n = 18) | |
| | A: 0.5 mg/kg (n = 18) | B: 0.2 mg/kg (n = 19) | | |
| *Demographics* | | | | |
| Sex, F:M | 16:2 | 16:3 | 13:5 | 0.232 |
| Age, yrs | 46.5 ± 11.6 | 48.1 ± 12.7 | 50.0 ± 7.6 | 0.340 |
| *Clinical profiles* | | | | |
| Disease duration, yrs | 12.3 ± 8.3 | 12.6 ± 9.3 | 11.5 ± 7.3 | 0.686 |
| Psychiatric comorbidities[a] | 2.0 ± 1.6 | 2.1 ± 1.4 | 2.4 ± 1.5 | 0.658 |
| Current antidepressants[b] | 1.7 ± 0.7 | 1.8 ± 0.8 | 1.5 ± 0.7 | 0.718 |

| HDRS-17 Scores | Groups A and B (n = 37) | | | Group C (x) (n = 18) | |
|---|---|---|---|---|---|
| Time points | Responders (x) | | Non-responders | | |
| **0 min (baseline)** | 22.4 ± 4.3 (x = 0) | [$p^1$= 0.485; $p^2$ = 0.556] | 24.8 ± 4.2 | 23.7 ± 4.5 (x = 0) | 0.547 |
| 40 min after treatment | 13.6 ± 4.2 (x = 9) | [$p^1$= 0.363; $p^2$ = 0.335] | 20.2 ± 4.6 | 20.9 ± 4.8 (x = 0) | 0.359 |
| 120 min after treatment | 11.7 ± 4.5 (x = 11) | [$p^1$= 0.043; $p^2$ = 0.048]* | 21.3 ± 4.9 | 20.1 ± 5.4 (x = 0) | **0.047** |
| **240 min after treatment** | 10.6 ± 3.7 (**x = 16**) | [$p^1$= 0.037; $p^2$ = 0.041]* | 20.8 ± 5.3 | 19.2 ± 6.1 (x = 2) | **0.040** |
| Day 2 (24 hours) after treatment | 10.1 ± 3.9 (x = 9) | [$p^1$= 0.043; $p^2$ = 0.046]* | 19.2 ± 5.7 | 18.9 ± 5.9 (x = 1) | **0.046** |
| Day 3 after treatment | 10.3 ± 4.4 (x = 9) | [$p^1$= 0.051; $p^2$ = 0.052] | 19.4 ± 6.3 | 19.1 ± 6.1 (x = 1) | 0.052 |
| Day 5 after treatment | 10.7 ± 4.6 (x = 5) | [$p^1$= 0.204; $p^2$ = 0.198] | 19.2 ± 6.2 | 19.2 ± 5.9 (x = 0) | 0.203 |
| Day 7 after treatment | 12.5 ± 5.0 (x = 3) | [$p^1$= 0.438; $p^2$ = 0.429] | 19.7 ± 6.1 | 19.7 ± 5.8 (x = 0) | 0.430 |

*x*: Number of subjects who showed a response to ketamine.
[a] Psychiatric comorbidities: chronic dysthymia, generalized anxiety disorder, panic disorder, social phobia and post-traumatic stress disorder.
[b] Number of antidepressants used, including regular antidepressants, mood stabilizers and atypical antipsychotics.
* $p < 0.05$ using Bonferroni-adjusted Tukey's post hoc test (responders vs. non-responders or the NS group). $p^1$ indicates the Bonferroni-adjusted p-values describing the HDRS-17 score differences between the responders and non-responders, and $p^2$ indicates the Bonferroni-adjusted p-values describing the HDRS-17 score differences between the responders and the NS group.

different ketamine responses.

a. EEG Power

The processed time-series EEG data were transformed into the frequency domain with a 256-point fast Fourier transform using Welch's method. Specifically, 10-min spans of data were analyzed with a 256-point moving window and a 128-point overlap. To increase the number of frequency bins, zero-padding was used to extend the length of the window, which can decrease the frequency interval and allows the power spectrum to appear smooth. Then, the windowed data were extended to 512 points by zero padding to calculate the power spectra, yielding an estimation of the power spectra from 1 to 12 Hz (frequency resolution: 0.5 Hz). The power spectra ($P$) of these windows were averaged and converted to a logarithmic scale. The relative EEG power was defined as the percentage of absolute power in any frequency band compared with the total power of the entire EEG spectrum.

The absolute power was marked as $P_{c,f}$, and the relative power was expressed as

$$\widehat{P_{c,f}} = P_{c,f}/P_{c,f'} \qquad (1)$$

where $c$ is the channel, and $f$ and $f'$ are the specific frequency band and all bands, respectively.

Notably, the absolute and relative EEG powers of the four prefrontal channels were calculated from the delta (1–3.5 Hz), theta (4–7.5 Hz), lower alpha (8–10 Hz) and upper alpha (10.5–12 Hz) bands.

b. EEG Alpha Asymmetry

We used the mid-prefrontal (Fp1/Fp2) and mid-lateral (AF7/AF8) hemispheric asymmetry index to establish a relative measure of the difference in EEG (lower and upper) alpha power between the right and left forehead areas. The formula used to calculate the left-right asymmetry score was defined as

$$A_{LR} = |(P_L - P_R)/(P_L + P_R)| \qquad (2)$$

where $P_L$ is the (lower/upper) alpha power that is left-lateralized at the baseline or post-treatment observations, and $P_R$ is the corresponding right-lateralized power.

c. EEG Theta Cordance

Theta cordance combines information from both the absolute and relative powers in the EEG theta band, which is less influenced by age, gender and severity associated with baseline depression.

The absolute and relative EEG powers from Equation (1) were normalized and are expressed as follows:

$$\tilde{P}_{A(c,f)} = P_{c,f}/max(P_c) \qquad (3)$$

$$\tilde{P}_{R(c,f)} = \hat{P}_{c,f}/max(\hat{P}_c) \qquad (4)$$

where $max(P_c)$ and $max(\hat{P}_c)$ represent the maximum absolute power and maximum relative power, respectively.

Then, theta cordance was calculated as follows:

$$Cordance_{c,f} = (\tilde{P}_{A(c,f)} - 0.5) + (\tilde{P}_{R(c,f)} - 0.5) \qquad (5)$$

D. EEG Predictors

The significant features from the EEG power analyses were treated as characteristic inputs to build a classification (prediction) model to distinguish ketamine effects at baseline. We considered a three-fold cross validation (66.7% training set and 33.3% testing set) to evaluate the classification performance of the mixed ketamine groups. Then, considering the prediction performances of ketamine groups A and B separately, we used the leave-one-subject-out cross-validation method to evaluate the classification performances of group A or B. The classification model was built based on EEG features from the training set and then examined via the testing set. To predict ketamine effects, we classified the responders and non-responders using EEG features at baseline. In this study, we employed various machine learning classifiers to identify the EEG features that could be used for comparison; these classifiers included linear discriminant analysis (LDA), nearest mean classifier (NMSC), k-nearest neighbors (k-NN with k = 3), Parzen density estimation (PARZEN), perceptron classifier (PERLC), discriminative restricted Boltzmann machine (DRBMC) and support vector machine with radial basis function (SVMRBF) (PRTools, http://prtools.org/; LIBSVM toolbox, http://www.csie.ntu.edu.tw/~cjlin/libsvm/). In addition, we employed the assessment criteria of medical screening (sensitivity and specificity) and pattern recognition (recall, precision and F-measure) to evaluate the performance of the proposed predictors.

### E. Statistical Analysis

We analyzed the differences in demographics, clinical profiles and treatment responses among the groups (A: 0.5 mg/kg ketamine, B: 0.2 mg/kg ketamine and C: NS) using one-way ANOVA tests, followed by correction of multiple comparisons by a sharpened Bonferroni procedure [34]. To investigate the differences in the EEG power between the responders and non-responders at baseline, we compared the EEG power values of the groups using the Wilcoxon rank-sum test to determine whether the two independent samples (sample size < 30) of responders and non-responders had the same distribution. To determine the EEG changes between the baseline and post-infusion time points, we compared the pre- and post-infusion EEG power and asymmetry values using the Wilcoxon signed-rank test. The significance level was set to $p < 0.05$. Additionally, since there are many potential EEG parameters to examine regarding ketamine responses, the experiment-wise error may be significantly inflated. Given the number of comparisons analyzed, Hochberg's sharpened Bonferroni [34] adjusted significant values (primary significance level $p < 0.05$ and secondary significance level $p < 0.025$) were used to reduce Type I error. Notably, the significantly different EEG baseline power ($p < 0.05$) between the responders and non-responders was treated as a characteristic input to build the classification (prediction) model. The statistical analyses were performed using the SPSS software package (version 15.0) and MATLAB (2011a) Bioinformatics Toolbox.

## III. RESULTS

### A. Demographic, Clinical and Treatment-Response Characteristics

Fifty-five outpatients with TRD were enrolled in this study between October 2014 and April 2016, including 18,

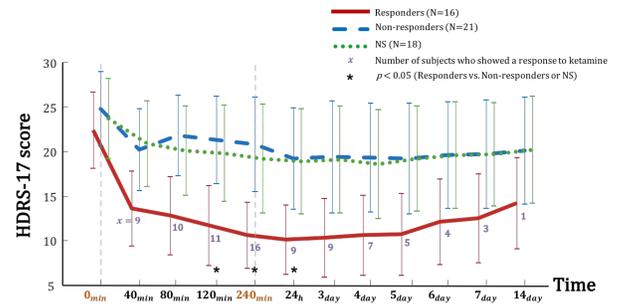

Figure 2. Changes in HDRS-17 scores from baseline to 14 days after infusion.

19 and 18 patients who received infusions of 0.5 mg/kg ketamine (group A), 0.2 mg/kg ketamine (group B) and normal saline (NS; group C), respectively. As shown in Table I, the demographic and clinical characteristics, including age, gender and clinical profiles (disease duration, psychiatric comorbidities and number of current antidepressants), were similar among the three groups. The concomitant medication regimens consisted of one antidepressant (36.4%), more than 2 antidepressants (21.8%) and a combination of antidepressants, mood stabilizers and atypical antipsychotics (41.8%). According to the MINI diagnosis, the most common psychiatric comorbidities were chronic dysthymia (52.7%), generalized anxiety disorder (GAD, 49.1%), panic disorder (32.7%), social phobia (10.9%) and post-traumatic stress disorder (10.9%). Approximately 54.5% of the patients had 2 or more comorbid psychiatric disorders. No significant differences in the prevalence of concomitant medications or comorbid psychiatric diseases were observed among the three groups.

As shown in Fig. 2, we periodically recorded the changes in the HDRS-17 item scores of depression severity from baseline (pre-treatment) to 14 days post-treatment throughout the study. Notably, the ketamine responders showed a substantially greater improvement. Specifically, we did not observe a significant difference among groups A, B and C at baseline, but the responders had significantly lower HDRS scores than the non-responders or NS subjects 120 min, 240 min, 24 hours (on day 2) and 3 days post-treatment. According to the HDRS-17 estimates shown in Table I, a maximum of 16 responders was observed in the two ketamine groups (A: 0.5 mg/kg, $n = 11$; B: 0.2 mg/kg, $n = 5$) 240-min post-treatment. Specifically, 11 of the 18 (61%) patients infused with 0.5 mg/kg ketamine showed a significant response 240-min post-treatment, while only 5 of the 19 (26%) patients infused with 0.2 mg/kg ketamine showed a significant response at this time point.

Nevertheless, only a few subjects in the NS group ($n = 2$) displayed an effective response during the same period. The HDRS-17 estimates also revealed that the most effective treatment outcome among all time points occurred at 240-min post-ketamine infusion. Furthermore, the comparison of the HDRS-17 scores at baseline with those at 240-min post-treatment indicated that the severity of depression (mean ± standard deviation (SD)) in the ketamine responders decreased from 22.4 ± 4.3 to 10.6 ± 2.7 with a response rate of 52.7 ± 7.3%; the severity in the ketamine non-responders

**Table II. Relative EEG Power: Baseline Comparisons between the Ketamine Groups**

| Relative Power | Ketamine Groups (n = 37) | | | | | | |
|---|---|---|---|---|---|---|---|
| | A: 0.5 mg/kg dose (n = 18) | | | B: 0.2 mg/kg dose (n = 19) | | | $p^b$ |
| | Responders (n = 11) | Non-responders (n = 7) | p | Responders (n = 5) | Non-responders (n = 14) | p | |
| **AF7** | | | | | | | |
| Delta | 1.40 ± 0.26 | 1.34 ± 0.14 | N | 1.42 ± 0.18 | 1.33 ± 0.17 | N | N |
| Theta | 0.89 ± 0.06 | 0.95 ± 0.06 | **0.042** | 0.91 ± 0.05 | 0.92 ± 0.07 | N | N |
| Low Alpha | 0.81 ± 0.11 | 0.91 ± 0.09 | 0.045[a] | 0.83 ± 0.25 | 0.90 ± 0.26 | N | N |
| High Alpha | 0.76 ± 0.13 | 0.78 ± 0.08 | N | 0.79 ± 0.20 | 0.75 ± 0.14 | N | N |
| **Fp1** | | | | | | | |
| Delta | 1.41 ± 0.21 | 1.33 ± 0.12 | N | 1.41 ± 0.19 | 1.32 ± 0.16 | N | N |
| Theta | 0.89 ± 0.06 | 0.95 ± 0.06 | N | 0.93 ± 0.06 | 0.94 ± 0.07 | N | **0.038** |
| Low Alpha | 0.84 ± 0.10 | 0.90 ± 0.11 | N | 0.85 ± 0.20 | 0.99 ± 0.25 | N | N |
| High Alpha | 0.74 ± 0.09 | 0.77 ± 0.08 | N | 0.80 ± 0.20 | 0.73 ± 0.15 | N | N |
| **Fp2** | | | | | | | |
| Delta | 1.42 ± 0.20 | 1.28 ± 0.15 | N | 1.42 ± 0.20 | 1.28 ± 0.16 | N | N |
| Theta | 0.88 ± 0.03 | 0.96 ± 0.06 | **0.035** | 0.91 ± 0.04 | 0.95 ± 0.05 | **0.028** | **0.042** |
| Low Alpha | 0.81 ± 0.11 | 0.92 ± 0.07 | 0.039[a] | 0.83 ± 0.18 | 0.91 ± 0.27 | N | N |
| High Alpha | 0.74 ± 0.12 | 0.81 ± 0.09 | N | 0.79 ± 0.18 | 0.78 ± 0.17 | N | N |
| **AF8** | | | | | | | |
| Delta | 1.41 ± 0.26 | 1.32 ± 0.17 | N | 1.40 ± 0.21 | 1.30 ± 0.18 | N | N |
| Theta | 0.89 ± 0.06 | 0.95 ± 0.07 | N | 0.92 ± 0.07 | 0.93 ± 0.05 | N | N |
| Low Alpha | 0.82 ± 0.13 | 0.91 ± 0.06 | N | 0.84 ± 0.21 | 0.90 ± 0.26 | N | N |
| High Alpha | 0.75 ± 0.15 | 0.78 ± 0.12 | N | 0.80 ± 0.15 | 0.76 ± 0.16 | N | N |

Delta: (1-3.5 Hz) / (1-12 Hz) EEG power (dB).
Theta: (4-7.5 Hz) / (1-12 Hz) EEG power (dB).
Low Alpha: (8-10 Hz) / (1-12 Hz) EEG power (dB).
High Alpha: (10.5-12 Hz) / (1-12 Hz) EEG power (dB).
$p$-value: measured using the Wilcoxon rank-sum test with a significant $p$-value < 0.05.
  1) N: Not significant ($p \geq .05$).
  2) Significance values were adjusted using a sharpened Bonferroni correction for multiple comparisons (primary significance level $p < 0.05$ and secondary significance level $p < 0.025$).
[a] Not statistically significant after sharpened Bonferroni adjustment.
[b] Comparisons of the responders in groups A and B.

decreased from 24.8 ± 4.2 to 20.8 ± 5.3 with a response rate of 18.4 ± 11.6%; and the severity in the NS group decreased from 23.7 ± 4.5 to 19.2 ± 6.1 with a response rate of 20.1 ± 13.0%. Notably, the severity of depression in the responders showed a significantly greater improvement than that in the non-responders or NS subjects ($p < 0.05$).

### B. Baseline Comparisons of EEG Power

As shown in Fig. 3, we investigated baseline comparisons (pre-treatment) between the responders and non-responders by combing the continuous power spectra of all channels (AF7, Fp1, Fp2 and AF8). Compared with the non-responders, the responders in the 0.5 mg/kg ketamine group showed lower relative EEG theta and lower alpha power ($p < 0.05$). There were no significant findings in the 0.2 mg/kg ketamine group.

As indicated by the baseline comparisons of the mean power spectra shown in Table II, compared with the non-responders in group A (0.5 mg/kg ketamine), the responders in group A had significantly ($p < 0.05$) weaker relative EEG power in the theta band on the AF7 and Fp2 channels and a decreasing trend towards relatively low alpha EEG power on the AF7 and Fp2 channels. In the baseline comparisons in group B (0.2 mg/kg ketamine), the responders showed significantly ($p < 0.05$) weaker relative EEG power in the theta band on the Fp2 channel than the non-responders. No differences were observed in the relative EEG power in the delta and high alpha bands between the responders and non-responders in groups A or B. Additionally, the responders in group A showed significantly ($p < 0.05$) higher relative EEG power in the theta band on the Fp1 and Fp2 channels than the

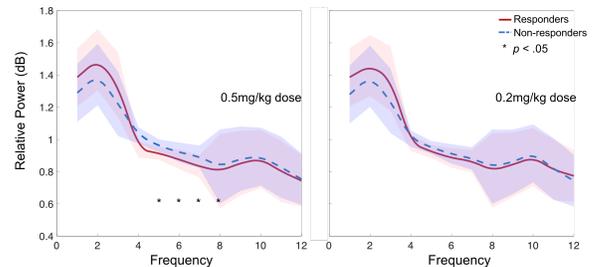

Figure 3. Baseline comparisons of the EEG power in the ketamine groups (left: 0.5 mg/kg, right: 0.2 mg/kg). Notably, the red and blue traces represent the mean ± SD of the EEG power in the responders and non-responders, respectively.

**Table III. Relative EEG Power, Alpha Asymmetry and Theta Cordance: Baseline vs. 240-min Post-treatment**

| | 0.5 or 0.2 mg/kg ketamine dose | | | | | | NS ($n = 18$) | | |
|---|---|---|---|---|---|---|---|---|---|
| | *Active group: Responders* ($n = 16$) | | | *Inactive group: Non-responders* ($n = 21$) | | | | | |
| Index | 0 min[b] | 240 min[c] | p | 0 min[b] | 240 min[c] | p | 0 min[b] | 240 min[c] | p |
| *Relative Power* | | | | | | | | | |
| **AF7** | | | | | | | | | |
| Delta | 1.40 ± 0.26 | 1.36 ± 0.26 | N | 1.31 ± 0.16 | 1.29 ± 0.21 | N | 1.16 ± 0.61 | 1.10 ± 0.57 | N |
| Theta | 0.90 ± 0.07 | 0.88 ± 0.08 | N | 0.95 ± 0.06 | 0.95 ± 0.13 | N | 0.91 ± 0.11 | 0.91 ± 0.11 | N |
| Low Alpha | 0.85 ± 0.15 | 0.96 ± 0.23 | **0.010** | 0.99 ± 0.25 | 0.97 ± 0.18 | N | 1.01 ± 0.27 | 1.08 ± 0.21 | N |
| High Alpha | 0.80 ± 0.19 | 0.85 ± 0.21 | N | 0.74 ± 0.12 | 0.78 ± 0.17 | N | 0.97 ± 0.58 | 1.01 ± 0.52 | N |
| **Fp1** | | | | | | | | | |
| Delta | 1.38 ± 0.21 | 1.31 ± 0.25 | N | 1.32 ± 0.12 | 1.29 ± 0.17 | N | 1.21 ± 0.56 | 1.18 ± 0.44 | N |
| Theta | 0.93 ± 0.08 | 0.91 ± 0.08 | N | 0.96 ± 0.06 | 0.97 ± 0.12 | N | 0.95 ± 0.11 | 0.93 ± 0.10 | N |
| Low Alpha | 0.87 ± 0.14 | 0.96 ± 0.18 | **0.023** | 0.98 ± 0.20 | 0.97 ± 0.18 | N | 0.95 ± 0.25 | 1.00 ± 0.17 | N |
| High Alpha | 0.78 ± 0.13 | 0.84 ± 0.19 | N | 0.73 ± 0.13 | 0.75 ± 0.16 | N | 0.88 ± 0.48 | 0.92 ± 0.43 | N |
| **Fp2** | | | | | | | | | |
| Delta | 1.41 ± 0.20 | 1.30 ± 0.22 | N | 1.26 ± 0.15 | 1.25 ± 0.19 | N | 1.10 ± 0.64 | 1.11 ± 0.50 | N |
| Theta | 0.90 ± 0.08 | 0.90 ± 0.09 | N | 0.97 ± 0.07 | 0.97 ± 0.12 | N | 0.96 ± 0.08 | 0.93 ± 0.22 | N |
| Low Alpha | 0.85 ± 0.15 | 0.98 ± 0.21 | **0.020** | 1.00 ± 0.22 | 1.00 ± 0.20 | N | 1.01 ± 0.29 | 1.06 ± 0.19 | N |
| High Alpha | 0.80 ± 0.13 | 0.87 ± 0.23 | **0.023** | 0.75 ± 0.16 | 0.79 ± 0.15 | N | 0.95 ± 0.57 | 0.98 ± 0.52 | N |
| **AF8** | | | | | | | | | |
| Delta | 1.39 ± 0.43 | 1.30 ± 0.20 | N | 1.30 ± 0.17 | 1.28 ± 0.19 | N | 1.59 ± 1.20 | 2.01 ± 5.63 | N |
| Theta | 0.92 ± 0.12 | 0.90 ± 0.08 | N | 0.95 ± 0.08 | 0.96 ± 0.11 | N | 0.95 ± 0.14 | 0.89 ± 0.20 | N |
| Low Alpha | 0.85 ± 0.16 | 0.98 ± 0.19 | **0.011**[a] | 0.99 ± 0.22 | 0.97 ± 0.19 | N | 0.82 ± 0.68 | 0.50 ± 0.12 | N |
| High Alpha | 0.80 ± 0.17 | 0.86 ± 0.15 | **0.045**[a] | 0.76 ± 0.12 | 0.79 ± 0.15 | N | 0.53 ± 1.14 | 0.36 ± 4.29 | N |
| *Alpha Asymmetry* | | | | | | | | | |
| **AF8 -AF7** | | | | | | | | | |
| Low Alpha | 0.12 ± 0.14 | 0.08 ± 0.12 | N | 0.06 ± 0.10 | 0.07 ± 0.09 | N | 0.06 ± 0.07 | 0.03 ± 0.03 | N |
| High Alpha | 0.11 ± 0.12 | 0.05 ± 0.04 | N | 0.10 ± 0.12 | 0.07 ± 0.08 | N | 0.07 ± 0.08 | 0.04 ± 0.04 | N |
| **Fp2 -Fp1** | | | | | | | | | |
| Low Alpha | 0.11 ± 0.12 | 0.03 ± 0.02 | **0.042** | 0.09 ± 0.11 | 0.06 ± 0.07 | N | 0.43 ± 1.48 | 0.09 ± 0.11 | N |
| High Alpha | 0.13 ± 0.18 | 0.07 ± 0.07 | N | 0.08 ± 0.10 | 0.10 ± 0.11 | N | 0.19 ± 0.43 | 0.10 ± 0.12 | N |
| *Theta Cordance* | | | | | | | | | |
| **AF7** | 1.55 ± 0.07 | 1.42 ± 0.08 | **0.041**[a] | 1.48 ± 0.08 | 1.39 ± 0.09 | N | 1.48 ± 0.08 | 1.41 ± 0.08 | N |
| **Fp1** | 1.54 ± 0.05 | 1.41 ± 0.06 | **0.014** | 1.50 ± 0.06 | 1.41 ± 0.07 | N | 1.51 ± 0.07 | 1.41 ± 0.09 | N |
| **Fp2** | 1.53 ± 0.05 | 1.42 ± 0.06 | **0.023** | 1.49 ± 0.07 | 1.40 ± 0.07 | N | 1.50 ± 0.06 | 1.40 ± 0.08 | N |
| **AF8** | 1.55 ± 0.08 | 1.41 ± 0.08 | **0.046**[a] | 1.47 ± 0.09 | 1.39 ± 0.08 | N | 1.48 ± 0.09 | 1.39 ± 0.09 | N |

Delta: (1-3.5 Hz) / (1-12 Hz) EEG power (dB).
Theta: (4-7.5 Hz) / (1-12 Hz) EEG power (dB).
Low Alpha: (8-10 Hz) / (1-12 Hz) EEG power (dB).
High Alpha: (10.5-12 Hz) / (1-12 Hz) EEG power (dB).
*p*-value: measured using Wilcoxon rank-sum test with a significant *p*-value < 0.05.
  1) N: Not significant ($p \geqslant 0.05$).
  2) Significance values were adjusted using sharpened Bonferroni correction for multiple comparisons (primary significance level $p < 0.05$ and secondary significance level $p < 0.025$).
[a] Not statistically significant after sharpened Bonferroni adjustment.
[b] Baseline (pre-treatment) period.
[c] 240-min post-treatment period.

responders in group B, indicating that for some patients, only a low ketamine dose (0.2 mg/kg) is needed to elicit a significant effect on EEG power.

### C. Comparisons of Baseline and 240 min Post-treatment

As shown in Table III, groups A and B were combined, and the responders and non-responders in this group were considered the active ketamine group and inactive ketamine group, respectively. We compared the relative EEG power before and after the ketamine or NS infusion (0 min vs. 240 min). The EEG power in the active ketamine group showed a significant ($p < 0.05$) increase in the low alpha band on the AF7, Fp1, Fp2 and AF8 channels 240-min post-treatment compared with that at baseline (pre-treatment), but no changes were observed in the delta or theta band. The relative EEG power in the inactive ketamine and placebo groups generally did not differ from the baseline values.

Table III also shows the differences in hemispheric asymmetry in the low and high alpha bands from the baseline to 240 min post-treatment observations. The midline (Fp1-Fp2) low alpha asymmetry in the active ketamine group showed a significant decrease ($p = 0.042$, from $0.11 \pm 0.12$ to $0.03 \pm 0.02$) after the ketamine treatment ($p < 0.05$), whereas the hemispheric alpha asymmetry in the inactive ketamine group and the NS group was similar between the baseline and post-treatment observations.

Additionally, we compared the EEG theta cordance from baseline to 240-min post-treatment (Table III). The responders showed significantly decreased EEG theta cordance in all channels ($p < 0.05$). Specifically, after the ketamine treatment, the responders showed a significantly decreased EEG theta cordance on the AF7 ($p = 0.041$, from $1.55 \pm 0.07$ to $1.42 \pm 0.08$), Fp1 ($p = 0.014$, from $1.54 \pm 0.05$ to $1.41 \pm 0.06$), Fp2 ($p = 0.023$, from $1.53 \pm 0.05$ to $1.42 \pm 0.06$) and AF8 ($p = 0.046$, from $1.55 \pm 0.08$ to $1.41 \pm 0.08$) channels, while the EEG theta cordance in the inactive ketamine group or the NS group showed no significant differences. Briefly, early changes in the EEG power, asymmetry and cordance were observed in the patients who actively responded to the ketamine infusion but not in those who received NS.

In summary, Table III presents the comparisons of baseline and 240 min post-treatment within one group (intra-subject EEG variabilities), which can identify ketamine responses and potentially contribute to a regression analysis in the future to estimate the relationships among EEG variables if long-term EEG recordings of TRD patients are conducted.

### D. Performances of EEG Predictors

The classification performances (mean accuracy ± SD) are summarized in Table IV. Specifically, we used seven machine learning classifiers (LDA, NMSC, 3-NN, PARZEN, PERLC, DRBMC and SVMRBF) to separately train a prediction model using two-thirds of the enrolled subjects ($n = 25$, including 11 responders and 14 non-responders) and then test the performance of the classifiers using the remaining subjects ($n = 12$). The significant EEG features from Table II (e.g., EEG power in relative theta or low alpha frequency band) were selected as input features.

In this study, we also considered the imbalanced problem; therefore, we randomly over-sampled 3 samples from the training dataset of the responder group to increase the sample of minority. Thus, the training dataset was balanced between the responder ($n = 14$) and non-responder ($n = 14$) groups. The EEG features combining the relative theta with the low alpha power had the best performance and the highest accuracy ($81.3 \pm 9.5\%$). The predictor with the best performance exhibited $82.1 \pm 8.6\%$ sensitivity, $91.9 \pm 7.4\%$ specificity, $81.9 \pm 8.6\%$ recall, $92.0 \pm 8.2\%$ precision and $57.2 \pm 4.4\%$ F-measure based on our SVMRBF (parameters: $c = 10$, $\gamma = 1$) predictor.

Then, we considered the classification performances of ketamine groups A and B separately. Regarding group A, we randomly over-sampled 2 samples from the training dataset of the non-responder group to increase the sample of minority. Thus, the training dataset was balanced between the responder ($n = 7$) and non-responder ($n = 7$) groups. The SVMRBF-based predictor (parameters: $c = 10$, $\gamma = 1$) achieved the highest accuracy ($80.6 \pm 8.3\%$), sensitivity ($81.7 \pm 8.5\%$), specificity ($88.3 \pm 7.5\%$), recall ($80.6 \pm 7.8\%$), precision ($91.1 \pm 7.7\%$) and F-measure ($57.3 \pm 4.5\%$). Regarding group B, we randomly over-sampled 6 samples from the training dataset of the responder group to increase the sample of minority. Thus, the training dataset was balanced between the responder ($n = 9$) and non-responder ($n = 9$) groups. The SVMRBF-based predictor (parameters: $c = 10$, $\gamma = 1$) achieved the highest accuracy ($78.4 \pm 9.6\%$), sensitivity ($79.3 \pm 8.8\%$), specificity ($84.2 \pm 7.7\%$), recall ($78.5 \pm 8.0\%$), precision ($87.0 \pm 7.9\%$) and F-measure ($52.6 \pm 5.5\%$).

When the other EEG baseline features (e.g., alpha asymmetry or theta cordance) were treated as characteristic inputs to build the classification (prediction) model, the

**Table IV. Performances of EEG Power Predictors**

| | *Accuracy ± SD by EEG Frequency (%)* | | | | | | | | |
|---|---|---|---|---|---|---|---|---|---|
| **Predictors** | *Mixed Ketamine Groups* | | | *0.5 mg/kg Ketamine Group* | | | *0.2 mg/kg Ketamine Group* | | |
| | *Theta* | *Low alpha* | *Theta + Low alpha* | *Theta* | *Low alpha* | *Theta + Low alpha* | *Theta* | *Low alpha* | *Theta + Low alpha* |
| *LDA* | 69.8 ± 6.5 | 60.2 ± 5.1 | 71.6 ± 7.4 | 70.2 ± 6.0 | 64.5 ± 6.0 | 72.1 ± 7.6 | 69.8 ± 6.5 | 60.2 ± 5.1 | 69.6 ± 8.7 |
| *NMSC* | 61.4 ± 5.4 | 62.5 ± 6.9 | 76.2 ± 9.3 | 65.3 ± 5.8 | 64.6 ± 6.3 | 78.3 ± 8.9 | 61.4 ± 5.4 | 62.5 ± 6.9 | 77.1 ± 9.0 |
| *3-NN* | 67.5 ± 6.5 | 61.1 ± 5.3 | 74.2 ± 9.7 | 68.7 ± 6.2 | 63.2 ± 5.8 | 75.2 ± 9.3 | 67.5 ± 6.5 | 61.1 ± 5.3 | 73.5 ± 9.1 |
| *PARZEN* | 67.8 ± 6.6 | 62.0 ± 5.8 | 75.9 ± 7.8 | 70.9 ± 6.1 | 63.5 ± 5.7 | 75.3 ± 8.0 | 67.8 ± 6.6 | 62.0 ± 5.9 | 74.1 ± 8.0 |
| *PERLC* | 60.6 ± 5.9 | 54.6 ± 5.4 | 70.4 ± 6.2 | 64.2 ± 6.5 | 59.4 ± 5.8 | 71.6 ± 5.9 | 60.6 ± 5.9 | 55.3 ± 6.0 | 69.7 ± 6.0 |
| *DRBMC* | 61.5 ± 6.5 | 57.3 ± 7.4 | 71.1 ± 4.7 | 67.3 ± 6.4 | 55.8 ± 7.2 | 73.0 ± 5.1 | 61.5 ± 6.5 | 56.4 ± 7.5 | 70.3 ± 5.2 |
| *SVMRBF* | 73.5 ± 9.5 | 63.1 ± 9.8 | **81.3 ± 9.5** | 77.6 ± 8.6 | 69.3 ± 7.4 | **80.6 ± 8.3** | 73.5 ± 9.5 | 64.3 ± 9.5 | **78.4 ± 9.6** |

Abbreviations: LDA: linear discriminant analysis; NMSC: nearest mean classifier; 3-NN: k-nearest neighbors with k = 3; PARZEN: Parzen density estimation; PERLC: perceptron classifier; DRBMC: discriminative restricted Boltzmann machine; SVMRBF: support vector machine with radial basis function.

performances were inferior to the EEG power predictors. Specifically, the EEG alpha asymmetry predictors could only achieve 73.1 ± 9.2% accuracy, and the EEG theta cordance predictors could only achieve 69.5 ± 8.6% accuracy. In summary, our proposed model could predict the antidepressant effects after the administration of ketamine (240 min) to distinguish the responders and non-responders at the baseline EEG power.

## IV. DISCUSSION

Our study investigated baseline and post-ketamine infusion forehead EEG patterns in patients with TRD under randomized, double-blind, placebo-controlled conditions. The rapid antidepressant effects of low-dose ketamine are reflected by changes in the EEG power spectra and asymmetry in the prefrontal region. We differentiated responders from non-responders who received 0.5 or 0.2 mg/kg ketamine infusions based on their baseline EEG and explored the EEG changes during the early (240 min) post-infusion period in the ketamine responders, ketamine non-responders and NS controls. According to Table III, the ketamine responders displayed weaker relative EEG power in the theta bands at baseline, increased relative EEG power and decreased asymmetry in the low alpha band; these responses were correlated with the rapid antidepressant effects of ketamine. Notably, our study used machine learning technology (Table IV) and wearable EEG devices to distinguish the ketamine responders from non-responders with 81.3 ± 9.5% accuracy, 82.1 ± 8.6% sensitivity and 91.9 ± 7.4% specificity; this method may have the potential to predict rapid antidepressant effects. Our forehead EEG patterns suggested that the baseline and early changes in prefrontal activities may predict individual antidepressant responses to a mixed dose of ketamine.

The dose-related efficacy of ketamine in the current study is consistent with an earlier meta-analysis that suggested that higher ketamine doses were more effective across several studies [35]. Moreover, one study investigating the S-isomer of ketamine suggested that both the 0.2 and 0.4 mg/kg doses were effective [36]. Nevertheless, R/S-ketamine has routinely been infused in previous studies at a dose of 0.5 mg/kg to treat depression [27, 37] but has never been tested at a lower dose (e.g., 0.2 mg/kg). Therefore, testing a lower R/S-ketamine dose to determine whether the antidepressant efficacy could be maintained while minimizing adverse psychoactive effects, such as dissociation, could be of interest. Furthermore, according to Table I, our findings show that 2 patients in group C responded to NS, which is the placebo effect in clinical trials. Prior studies have shown that depressed patients who are treated with placebo may exhibit substantial reductions in symptoms [38]. The placebo effect involves psychological factors, including frequent visits, the infusion procedure, care from medical staff members, and some brain mechanisms of the placebo effect. Additionally, of particular interest, the efficacy of ketamine in Han Chinese, e.g., Taiwanese, patients [39] has been noted, but the generalizability of ketamine efficacy to populations of diverse races and ethnicities has not been established.

Therefore, our study addresses this need by examining the effect of ketamine in a Taiwanese population. Our finding that Taiwanese patients may experience good antidepressant effects with a 61% response rate to the standard (0.5 mg/kg) ketamine dose is consistent with our recent adjunctive ketamine study [39].

### A. Baseline EEG Patterns

Baseline measures of prefrontal electrical activity have been highlighted as EEG biosignatures of antidepressant responses. Theta and low alpha activity have been linked to processing functions related to emotion [40], and pre-treatment theta activity has been associated with the antidepressant response localized to the anterior cingulate cortex (ACC) [41]. According to our baseline results, as shown in Table II, the patients with TRD who exhibited an effective ketamine response exhibited lower relative EEG power in the theta and low alpha bands, whereas the absolute EEG power differences were not significant in any of the frequency bands. Moreover, according to our results, as described in section III-A, the 0.5 mg/kg dose of ketamine produced better response effects than the 0.2 mg/kg dose (response rate = 61.1% vs. 26.3%) 4 hours after treatment, which may be explained by the fact that most patients responded to the 0.5 mg/kg dose, whereas the lower dose (0.2 mg/kg) affected only some patients with weak EEG power in the theta band at baseline. Thus, patients who have relatively weak theta waves at baseline may require a normal low dose (0.5 mg/kg) of ketamine; otherwise, they could require a slightly lower dose (0.2 mg/kg). Our baseline EEG power results are consistent with a previous study using selective serotonin reuptake inhibitor (SSRI) treatment [17] but partially contradict a study that associated fluoxetine and venlafaxine antidepressant responses with overall higher baseline EEG theta power [16]. The differences in these findings may be related to the differences in the mechanisms of action of rapid antidepressants (e.g., ketamine) and conventional antidepressants. Regarding the biological mechanism derived from theta and alpha activity, lower baseline theta and alpha activity may be positively correlated with glucose metabolism in ketamine responders [42] and reflect lower arousal associated with lower serotonergic activity [18]. Additionally, the current interest in using baseline biosignatures to predict the treatment response in patients with depression has been reported [43, 44]. Prefrontal EEG power at baseline was reported to predict the SSRI antidepressant response with 63% [17] and 88% accuracy [45] in two studies using machine learning technology, suggesting that the use of forehead EEG patterns has potential for building a baseline predictor of the responses to ketamine treatment.

### B. Early Changes in Prefrontal EEG After Ketamine Treatment

EEG is generally used to evaluate antidepressant outcomes by examining pre- to post-treatment EEG changes [44]. Using baseline relative EEG measurements and early post-treatment changes, significantly lower alpha power has been observed in depressed patients compared with normal controls [46], and significantly higher alpha asymmetry has been observed in depressed patients compared with that in normal controls in previous baseline EEG studies [47]. These findings revealed EEG changes in the active ketamine group compared with the normal controls. Based on our findings, as shown in Table III, the early (240 min) changes in the post-treatment responders, which are characterized by increased relative power and decreased asymmetry in the low alpha band, were required for the ketamine-specific rapid antidepressant effect. Our results revealed increased prefrontal alpha activity after effective responses. This finding is consistent with a previous study showing that elevated resting rACC activity is correlated with the ATR through adaptive self-referential functions, such as mindfulness and non-evaluative self-focus [48]. The decreased EEG asymmetry may be positively correlated with lower behavioral activation sensitivity and inversely correlated with negative affect and behavioral inhibition [47, 49]. Regarding EEG cordance, our findings showed a decrease in theta cordance at prefrontal leads compared to that at baseline; however, this finding was not observed in the non-responders or patients treated with placebo. This trend is similar to that reported in prior studies using other antidepressant treatments [24, 25].

However, although correlations between these EEG signatures and the response to antidepressant treatment have been proposed, none have been validated to date because recent findings have not been consistently replicated. Jaworska, et al. [50] reported an elevated absolute alpha power in depressed patients. This finding was replicated in a large sample with increased alpha and theta EEG power among patients during the early stages of depression [51]. Other studies failed to show decreased alpha activity in patients with TRD [46]. In addition, the EEG differences have been associated with improved antidepressant treatment outcomes. For example, an increased EEG alpha power has been shown to differentiate responders from non-responders following 3 to 6 weeks of treatment with SSRIs, such as paroxetine and fluoxetine [21]. Nevertheless, many studies have failed to replicate the alpha asymmetry findings in depression [20, 46]. In addition to alpha activity measures, one prognostic study reported that reduced theta power was related to the response to antidepressant treatments [17]. In contrast, increased theta activity has been observed in responders 6 weeks after treatment with SSRIs [52]. These studies are limited due to the mixed results of theta and alpha activity associated with different antidepressant responses.

*C. Ketamine Outcome Prediction*

Current research has focused on using baseline patterns to predict treatment responses in depression [43, 44]. Abnormalities in theta and alpha EEG activities have been associated with major depression [11, 12, 15, 16, 18], suggesting that the EEG parameters of the theta and alpha bands could serve as significant biosignatures and offer the potential to build a baseline predictor. A prior study reported that the baseline prefrontal relative theta power could predict the antidepressant response to SSRIs with 63% accuracy [17]. Another more recent study reported that baseline prediction improved to 87.9% accuracy for SSRI treatment using machine-learning technology and various patterns [45]. The baseline results of our study (Table IV) provided patterns of relative theta and low alpha EEG power that predict the effect of ketamine with $81.3 \pm 9.5\%$ accuracy, $82.1 \pm 8.6\%$ sensitivity and $91.9 \pm 7.4\%$ specificity based on the SVMRBF predictor. This baseline EEG predictor may offer a supporting reference to help doctors select an appropriate ketamine dose for depressed patients.

In contrast to traditional antidepressants, ketamine treatment can rapidly reduce depressive symptoms within 4 hours post-treatment. Considering the rapid effects of ketamine observed in our manuscript, we labeled the responders and non-responders 4 hours post-treatment. If we labeled the responders and non-responders at other time instances, e.g., after 4 hours post-treatment, the numbers of responders would decease instantly due to the reduced effects of ketamine; therefore, this methodology is not considered an optimized label and training model, accounting for the unbalanced samples after 4 hours post-treatment (e.g., increasing the numbers of responders and decreasing the numbers of non-responders).

*D. Wearable Forehead EEG with Dry Sensors*

Notably, the identified EEG signatures have the potential to distinguish ketamine responders and non-responders at baseline and predict the levels of antidepressant responses to ketamine with further clinical usage. Due to the development of sensor technology, an alternative to conventional EEG devices with wet electrodes and cables has emerged. Wearable and wireless EEG devices with dry-contact sensors (Fig. 1-A) have led to a reduction in the amount of preparatory work required for long-term monitoring and daily use [13, 14, 53] and have successfully been used to monitor brain activity in sleep and driving experiments [54, 55]. Wearable wireless EEG devices with dry sensors offer a promising tool for the daily monitoring of depression. Forehead EEG patterns could potentially be used to estimate ketamine responses in daily life via a wearable EEG-based system.

*E. Limitations*

The major strength of this study was the sizeable number of patients tested using a randomized and placebo-controlled design. However, this study also had limitations. First, the responders were defined as subjects with at least a 45% reduction in baseline depression symptoms (HDRS-17) instead of the popular criterion of a 50% reduction as the sample size of the responders was too small to conduct statistical analyses. Furthermore, we combined the data from groups A and B in the analysis because of the small number of responders in group B. In the future, we could overcome this problem by increasing the total number of participants in each dose group to obtain a better comparison of the responders at each dose. Second, the recruitment in the study was limited to Asian patients with TRD, which may have yielded results that cannot be generalized to other

************

populations. Wider application would require a broader range of participants of different races. Third, none of our participants reported a significant response with traditional antidepressant treatments; therefore, the EEG dynamics were primarily influenced by the effect of ketamine. Nevertheless, EEG changes may be affected by the concomitant use of other antidepressant treatments. Fourth, the results of the present study provide evidence supporting the involvement of EEG power and asymmetry in the low alpha band on the 240-min antidepressant effects of a single-dose ketamine treatment in patients with TRD. However, further research is required to determine how the changes in EEG power and asymmetry are related to the antidepressant effects of ketamine at other time points (days 2, 3, 4, etc.). Finally, as our wearable headband-style EEG equipment is limited by the placement of the electrodes (prefrontal area), this device is not feasible for covering the entire brain. Thus, our findings represent only regional specificity for prefrontal EEG dynamics. We are unsure whether the effects of ketamine could also involve the EEG dynamics of other brain regions (e.g., parietal or occipital cortices). Due to this limitation of the electrode placements, we plan to use a multichannel EEG device to obtain a wider range of brain signals in the future.

V. CONCLUSION

Our randomized, double-blind, placebo-controlled study revealed the EEG patterns at baseline and in response to a mixed-dose ketamine treatment in patients with TRD. The results provided initial evidence that ketamine responders had weaker baseline EEG theta power, which functioned as an EEG-based predictor with 81.3 ± 9.5% accuracy, 82.1 ± 8.6% sensitivity and 91.9 ± 7.4% specificity. Moreover, an active ketamine response was accompanied by increased alpha power and decreased alpha asymmetry. In addition, Taiwanese patients may exhibit a good antidepressant response to the standard (0.5 mg/kg) ketamine dose. These insights provide important information about potential targets for the baseline identification of ketamine responders and the underpinnings of the early effects of ketamine on TRD.

ACKNOWLEDGEMENTS

We express our gratitude to all patients who kindly participated in this study. In addition, we thank all research assistants (Hui-Ju Wu and Chia-Min Huang), physicians, pharmacists and the nursing staff at the D020 Unit of Taipei Veterans General Hospital for their assistance during the study process; without them, this work would not have been possible. We also greatly appreciate Dr. Guang Cheng for providing excellent opinions regarding our work.

*************

*************